\documentclass[12pt,a4paper]{JHEP3}
\usepackage{amsmath,amssymb,bm}
\usepackage{latexsym}
\usepackage{graphicx}

\def\eqref#1{Eq.~(\ref{#1})}

\newcommand{\nn}{\nonumber}

\def\X5sp{{\rm X}_5}
\def\Y3sp{{\rm Y}_3}
\def\Z3sp{{\rm Z}_3}

\newcommand{\bea}{\begin{eqnarray}}
\newcommand{\eea}{\end{eqnarray}}

\def\X5sp{{\rm X}_5}
\def\Y3sp{{\rm Y}_3}
\def\Z3sp{{\rm Z}_3}

\title{Spectrum from the warped 
compactifications with the de Sitter universe}
\author{
Masato Minamitsuji\\
Yukawa Institute for Theoretical Physics,
Kyoto University,\\
Kyoto, 606-8502, Japan.\\
~~~E-mail: \email{masato$``$at$"$yukawa.kyoto-u.ac.jp}}
\author{
Kunihito Uzawa\\
Department of Physics, School of Science and Technology, \\
Kwansei Gakuin University, Sanda 669-1337, Japan.\\
~~~E-mail: \email{uzawa$``$at$"$yukawa.kyoto-u.ac.jp}}

\abstract{
We discuss the spectrum
of the tensor metric perturbations and 
the stability
of warped compactifications with the de Sitter spacetime
in the higher-dimensional gravity.
The spacetime structure is given 
in terms of the warped product of 
the non-compact 
direction,
the spherical internal dimensions
and the four-dimensional de Sitter spacetime.
To realize 
a finite bulk volume,
we construct the brane world model,
using the cut-copy-paste method.
Then, we compactify the spherical directions
on the brane.
In any case,
we show
the existence of the massless zero mode and
the mass gap of it 
with massive Kaluza-Klein modes.
Although the brane involves the spherical dimensions,
no light massive mode is excited.
We also investigate the scalar perturbations,
and show that the model is unstable due to the 
existence of a tachyonic bound state,
which seems to have the universal negative mass square,
irrespective of the number of spacetime dimensions. 
}

\begin{document}


\section{Introduction}

The realization of inflation and dark energy
is the challenging issue in cosmology and particle physics.
Since string theory is higher-dimensional,
one has expected that
such an accelerating universe may be realized
via dynamics of the extra-dimensional space.
Realization of an accelerating universe 
depends on ansatz of the spacetime metric
and fields.
Whether such a solution
can serve as a realistic model will require more
studies.

The traditional expectation has been 
that one obtains
the four-dimensional de Sitter universe 
as the exact solution of higher-dimensional gravity,
in particular in superstring- or M-theory.
One of the most successful constructions of the 
de Sitter universe in higher dimensions
is 
the warped compactification
in five dimensions discussed in Refs. \cite{nihei,kaloper}.
This
is 
the cosmological generalization of the famous
Randall-Sundrum model \cite{rs}. 
Recently,
higher dimensional  
warped compactifications with the de Sitter universe
 in the pure gravity 
have been obtained in Ref. \cite{neupane2}.
New solutions of warped compactifications
including the bulk matter
have also been derived in Ref. \cite{mu}. 
In these solutions, the spacetime 
is given by the warped product of 
the non-compact extra dimension, the internal spherical direction
and the four-dimensional de Sitter spacetime.
As the bulk volume diverges, 
we construct a 
codimension-one braneworld, 
by cutting the spacetime
at a certain place of it, 
and then gluing the remaining peace
to its copy at the same position.
By construction,
there is the $Z_2$-symmetry
with respect to the brane.
This is 
the ordinary cut-copy-paste method.
The difference from the five-dimensional case
is that 
the braneworld involves the spherical dimensions
as well as the de Sitter spacetime.
We then compactify 
the spherical dimensions to obtain the 
four-dimensional cosmology.
Such a way 
of construction of 
the braneworld model
from a higher-dimensional theory
is known as the Kaluza-Klein braneworld
in the literature \cite{slss}.
The junction condition requires
the positive brane tension.
The insertion of the braneworld is hence
equivalent to 
adding 
a positive potential energy 
to the effective theory. 
Though these solutions cannot be 
counterexamples of the NO-GO theorem \cite{mn}, 
they give us an interesting class of 
the cosmological braneworld models in 
spacetimes of
higher than six dimensions.
In this paper, we study the spectrum of the 
gravitational waves and the stability
in our model.

In order to investigate
whether these models are realistic, 
we have to 
see
the localizability of the massless zero mode
which could reproduce the four-dimensional physics
after compactifying the spherical dimensions.
The inflationary four-dimensional universe after 
the compactification
also should not suffer
excitations of any light massive mode.
Here, we define the light mode as follows:
The effective four-dimensional metric after the compactification
on the brane
is given by the de Sitter spacetime
\bea
ds_4^2=-dt^2+c_0^2 e^{2Ht}\delta_{ij}dx^i dx^j,
\label{dS}
\eea
where $H$ is the Hubble expansion rate
and $c_0$ is the 
size of the universe at $t=0$.
In the effective four-dimensional point of view,
the time evolution equation 
of each four-dimensional mode 
of mass $m$ and comoving momentum $k$
whose mode function is given by $\varphi_{m,k}$, 
is written as
\bea
\Big(\frac{d^2}{dt^2}+3H\frac{d}{dt}+\frac{k^2}{c_0^2e^{2Ht}}
+m^2\Big)\varphi_{m,k}(t)=0.
\label{4d_mode_eq}
\eea
The solution to Eq. (\ref{4d_mode_eq}) is given by
\bea
\varphi_{m,k} \propto
e^{-\frac{3}{2}H t}
Z_{i\nu_m}\Big(\frac{k}{c_0}e^{-Ht}\Big),\label{4dmode}
\quad 
\nu_m:=\frac{m^2}{H^2}-\frac{9}{4}=:\frac{1}{H^2}(m^2-m_c^2)\,,
\eea
where $Z_{\nu}$ denotes the Bessel functions
of order $\nu$.
Let us review the behavior in the homogeneous limit $k\to 0$.
The late time behavior of Eq. (\ref{4dmode})
depends on the mass.
Heavy modes of $m> m_c=\frac{3}{2}H$ decay rapidly as
$|\varphi_{m,k}|\propto a^{-\frac{3}{2}}$,
while light modes of $m<m_c$ decay more slowly and in particular for 
$m\ll m_c$, $|\varphi_{m,k}|\propto a^{-\frac{3}{4}\frac{m^2}{m_c^2}}$.
Thus the contribution of heavy KK modes is diluted rapidly during inflation,
while that of light ones may survive and affect the late time cosmology.
We will show 
that in any model the mass gap between the zero and 
massive modes
is always greater than $\frac{3}{2}H$,
and hence warped compactifications with the de Sitter universe
are free from massive excitations.
The five-dimensional model has been investigated
in e.g., Ref. \cite{langlois},
which showed that there is always mass gap given by $\frac{3}{2}H$.

This paper is constructed as follows.
In Sec. II, we review
warped compactification solutions with the de Sitter universe
 discussed 
in \cite{neupane2,mu}.
In Sec. III, we investigate the tensor perturbations
with respect to the three-dimensional space
and mass spectrum.
In Sec. IV, we discuss the stability
of the solutions with a spherical internal space. 
The last Sec. V
is devoted to give summary and conclusion.

\section{Warped compactifications with the de Sitter spacetime}

In this section, we review solutions of
warped compactifications with the de Sitter spacetime
 discussed in \cite{neupane2}
as well as \cite{mu}.

\subsection{The five-dimensional model}

We consider the Einstein gravity with a negative cosmological constant
in a five-dimensional spacetime
including the braneworld 
\bea
S=\frac{1}{2\kappa_5^2}\int d^5 x\sqrt{-g}\Big(R-2\Lambda_5\Big)
-\int d^4x\sqrt{-q}\sigma_5,
\eea
where $\Lambda_5<0$ is 
the
cosmological constant,
$\sigma_5$ is the brane tension and $q_{\mu\nu}$
is the induced metric on the brane.
There is the warped compactification solution with the de Sitter universe
\bea
ds^2
=A(y)^2
\Big(
-dt^2
+c_0^2 e^{2Ht}\delta_{ij}dx^i dx^j
+dy^2
\Big),\label{5d}
\eea
where 
\bea
A(y)^{-1}=e^{H|y|}
-\frac{|\Lambda_5| e^{-H|y|}}
      {24H^2}.
\eea
The bulk solution 
is
the five-dimensional anti de Sitter (AdS) spacetime 
\cite{nihei,kaloper,langlois}.
By defining the curvature radius of the AdS spacetime
 $\ell:=\sqrt{\frac{6}{|\Lambda_5|}}$,
the warp factor $A$ is rewritten
into the more familiar form
\bea
A(y)^{-1}
=\frac{1}{H\ell}
\sinh \big[H(|y|+y_0)\big],\quad
e^{Hy_0}:=2(H\ell).
\eea 
To realize the finite volume of the extra space,
the brane boundary is put at $y=y_b>0$
and the $Z_2$-symmetry 
across it
is also
imposed.
The brane position $y=y_b$
is given by
\bea
e^{H y_b}=\frac{1}{2}\Big(1+\sqrt{1+\frac{|\Lambda_5|}{6H^2}}\Big).
\eea
The induced metric on the brane 
is that of
the four-dimensional de Sitter spacetime Eq. (\ref{dS}).
The junction condition gives $\kappa_5^2 \sigma_5
=\sqrt{6(6H^2+|\Lambda_5|)}>0$.

\subsection{The $D(>6)$-dimensional models}

In this subsection, 
we consider warped compactification solutions
of $D(>6)$ dimensions
with a spherical internal space.

\vspace{0.2cm}

\underline{\it (1) In the case of the pure gravity}

\vspace{0.2cm}


First of all, we consider the $D$-dimensional Einstein gravity
including the braneworld
\bea
S=\frac{1}{2\kappa_D^2}\int d^{D} x \sqrt{-g}R
-\int d^{D-1}x\sqrt{-q}\sigma_D,
\eea
where $\sigma_D$ is the brane tension and $q_{\mu\nu}$
is the induced metric on the brane.
The warped compactification of the de Sitter universe
\cite{neupane2}
is given by
\bea
ds^2
=
A(y)^2
\Big[
-dt^2
+c_0^2 e^{2H t}\delta_{ij}dx^i dx^j
+\frac{1}{H^2}
\Big(G(y) dy^2+ \frac{D-6}{3}d\Omega_{D-5}^2\Big)
\Big], \label{nd0}
\eea
with 
\bea
A(y)=\Big(\frac{\cosh\big[M(|y|+a)\big]}{\cosh (Ma)}\Big)^{-p},\quad
G(y)= \frac{D-2}{3}p^2 M^2 \tanh^2 \big[M(|y|+a)\big].
\eea
We assume $p>0$ and
redefine 
the coordinate 
of the noncompact direction 
$dy=\frac{H}{G^{\frac{1}{2}}}dY$,
which gives 
\bea
Y=\frac{p}{H}\sqrt{\frac{D-2}{3}}\ln \cosh [M(|y|+a)].
   \label{y}
\eea
For $p>0$, the direction of the increasing $y$ corresponds to
that of the increasing $Y$.
Setting the coordinate $Y$, the $D$-dimensional metric (\ref{nd0})  
is rewritten by
\bea
ds^2
=
A(Y)^2
\Big[
-dt^2
+c_0^2 e^{2H t}\delta_{ij}dx^i dx^j
+dY^2
+ \frac{D-6}{3H^2}d\Omega_{D-5}^2
\Big], \label{nd}
\eea
where the warp factor $A$ is 
\bea
\label{ndnd}
A(Y)=e^{-H\sqrt{\frac{3}{D-2}}|Y-Y_0|}\,.
\eea

The spacetime structure
is given in terms of the warped product of 
$dS_4$, $\mathbb{R}$
and $S^{D-5}$.
We emphasize that in the new frame $Y$
there is no explicit dependence of the metric on $p$ and $M$,
which indicates that 
these parameters are not physical
and can be absorbed by the redefinition of coordinates. 
We assume that the $(D-1)$-dimensional braneworld
is located at
$Y=Y_0=\frac{p}{H}\sqrt{\frac{D-2}{3}}\ln \cosh (Ma)$
where $A(Y_0)=1$.
And we impose the $Z_2$-symmetry across it.
The conformal metric
in the square bracket of \eqref{nd}
is given by the product
of the four-dimensional de Sitter spacetime,
the noncompact 
$Y$ direction
and the $(D-5)$-dimensional sphere.
The internal space has a deficit solid angle
at $Y\to\infty$,
given by 
\bea
\Delta \Omega^{(D)}_{D-5}
=\Omega_{D-5}
\Big[
1
-\Big(
  \frac{D-6}{D-2}\Big)^{\frac{D-5}{2}}
\Big].
\eea
The induced metric on the brane is given by 
\bea
\label{oku}
ds_{\rm ind}^2
=
-dt^2
+c_0^2 e^{2Ht}\delta_{ij}dx^i dx^j
+\frac{D-6}{3H^2}d\Omega_{D-5}^2.
\eea
The junction condition gives $\kappa_D^2\sigma_D=2\sqrt{3(D-2)}H$.

\vspace{0.2cm}

We now generalize the solution Eq. (\ref{nd})
to the case with a cosmological constant
\bea
S=\frac{1}{2\kappa_D^2}\int d^{D} x \sqrt{-g}
\Big(R-2\Lambda_D\Big)
-\int d^{D-1}x\sqrt{-q}\sigma_D.
\eea
The solution discussed in \cite{neupane2}
is given by 
the metric Eq. (\ref{nd0})
with 
\bea
&&A(y)=\Big(\frac{\cosh\big[M(|y|+a)\big]}{\cosh (Ma)}\Big)^{-1},\quad
\nn
\\
&&G(y)= \frac{(D-2) M^2 \sinh^2 \big[M(|y|+a)\big]}
{3\cosh^2\big[M(|y|+a)\big]
+\frac{2|\Lambda_D|}{(D-1)H^2}\cosh^2(Ma)}.
\label{ds_scale}
\eea
This solution is
the generalization
of that in the previous subsection with $p=1$.
As before,
it is also useful to redefine the coordinate 
$dy=\frac{H}{G^{\frac{1}{2}}}dY$,
which gives 
\bea
Y&=&\frac{1}{H}\sqrt{\frac{D-2}{3}}
\ln\Big[\frac{1}{2}
\Big( \cosh [M(|y|+a)]
\nn\\
&+&\sqrt{\cosh^2[M(|y|+a)]+\frac{2|\Lambda_D| \cosh^2(M a)}{3(D-1)H^2}}
\Big)
\Big].
\eea
The spacetime metric 
can be rewritten as Eq. (\ref{nd})
with
\bea
\label{cosmo}
A(Y)&=&\Big(
 \cosh\big[\sqrt{\frac{3}{D-2}}H|Y-Y_0|\big]
\nn\\
&+&\sqrt{1+\frac{2|\Lambda_D|}{3(D-1)H^2}}
 \sinh \big[\sqrt{\frac{3}{D-2}}H|Y-Y_0|\big]
\Big)^{-1}.
\eea
The internal space has a deficit solid angle
at $Y\to \infty$
,
given by 
\bea
\Delta \Omega^{(D)}_{D-5}
=\Omega_{D-5}
\Big[
1
-\Big(
  \frac{D-6}{D-2}\Big)^{\frac{D-5}{2}}
\Big].
\eea
The $(D-1)$-dimensional braneworld is located at 
$$Y=Y_0:=\frac{1}{H}
\sqrt{\frac{D-2}{3}}
\ln\Big[\frac{\cosh(M a)}{2}
\Big(1+\sqrt{1+\frac{2|\Lambda_D|}{3(D-1)H^2}}\Big)
\Big].$$
We impose the $Z_2$-symmetry in the $Y$-direction across it.
The induced metric on the brane is 
the same as Eq. (\ref{oku}).
The junction condition gives
$\kappa_D^2\sigma_D
=2H\sqrt{3(D-2)\big(1+\frac{2|\Lambda_D|}{3(D-1)H^2}\big)}$.

\vspace{0.2cm}

\underline{\it (2) In the case of the scalar-tensor theory}

\vspace{0.2cm}

We consider the scalar-tensor theory with a negative potential
including the braneworld
\bea
S=\frac{1}{2\kappa_D^2}
\int d^{D}x\sqrt{-g}
\Big(
R
-\frac{1}{2}\big(\partial\phi\big)^2
-2e^{\beta \phi}\Lambda_s
\Big)
-\int d^{D-1}x\sqrt{-q}\sigma_s e^{\gamma_s \phi},
\eea
where $\beta$ represents the coupling parameter,
$\Lambda_s<0$ is constant,
and $\gamma_s$ denotes the brane coupling to the scalar field.
The solution discussed in \cite{mu}
is given 
by the metric Eq. (\ref{nd}) with
\bea
A=e^{-H\sqrt{\frac{3(\frac{2}{D-2}+c)}{c(D-2)}}|Y-Y_0|}.
\eea
The scalar field configuration is given by 
\bea
\phi
&=&2H \sqrt{\frac{3}{c(D-2)}}|Y-Y_0|,
\label{sol_scalar}
\quad
c:=\beta^2-\frac{2}{D-2}.
\eea
The expansion rate of the de Sitter universe is given by
\bea
H^2=-\frac{1}{3}c\Lambda_s.
  \label{def_Hubble}
\eea
According to $\Lambda_s<0$, one has $\beta^2>\frac{2}{D-2}$ 
so that $H^2>0$.
The internal space has a deficit or surplus
solid angle 
at $Y\to \infty$
which is given by 
\bea
\Delta \Omega^{(s)}_{D-5}
=\Omega_{D-5}
\Big[
1
-\Big\{
  \frac{D-6}{D-2}\Big(1+\frac{2}{c(D-2)}\Big)
 \Big\}^{\frac{D-5}{2}}
\Big].
\label{def_scalar}
\eea
For $c=c_{\ast}:=\frac{D-6}{2(D-2)}$,
the internal space becomes a flat space,
where $\beta_{\ast}=\frac{1}{\sqrt{2}}$.
The metric (\ref{def_scalar}) gives a deficit  
solid angle for $c>c_{\ast}$ while 
the $D$-dimensional spacetime has a surplus solid angle for $c<c_{\ast}$.
We assume that the braneworld is located at $Y=Y_0$,
and we impose the $Z_2$ symmetry.
The induced metric on the brane is 
the same as Eq. (\ref{oku}).
The junction conditions of the metric and the scalar field
give
$\kappa_D^2\sigma_s=2\sqrt{\frac{3((D-2)c+2)}{c}}H$
and
$\gamma_s
=\frac{1}{(D-6)\beta}$.

\vspace{0.2cm}

\underline{\it (3) In the case with a form field strength}

\vspace{0.2cm}

Finally, we consider 
the theory with a form field strength including the braneworld 
\bea
S&=&\frac{1}{2\kappa_D^2}
\int  d^D x\sqrt{-g}
\Big[
R-2e^{-\frac{\alpha \phi}{D-6}}\Lambda_f
-\frac{1}{2}\big(\partial\phi\big)^2
-\frac{1}{2(D-5)!}e^{\alpha\phi}F_{(D-5)}^2
\Big]
\nonumber\\
&-&\int d^{D-1}x\sqrt{-q}\sigma_f e^{\gamma_f \phi}.
\label{fins}
\eea
With
the warp factor $A(Y)$,
the solution discussed in \cite{mu} 
is given by
\bea
&&ds^2=A(Y)^2
\Big[
-dt^2+c_0^2e^{2H t}\delta_{ij}dx^i dx^j
+dY^2
+\frac{D-6}{3H^2+\frac{f^2}{2}}
d\Omega_{D-5}^2
\Big],\label{flux}
\nonumber\\
&&
\phi=\frac{2(D-6)}{\alpha}\ln A,
\nonumber\\
&&
F=f
\Big(\frac{D-6}{3H^2+\frac{f^2}{2}}\Big)^{\frac{D-5}{2}}
\sqrt{\gamma}
 dz^1\wedge dz^2\wedge \cdots \wedge dz^{D-5},
\eea
where $\gamma_{ab}$ is the metric of an unit $S^{D-5}$,
$f$ represents the strength of form field,
and 
the warp factor
and the expansion rate are given by 
\bea
 A(Y)=e^{-\sqrt{\frac{3}{D-2}\Big(1+\frac{D-6}{(D-2)\zeta}\Big)} 
H|Y-Y_0|}\,,~~~~
H^2=\frac{\zeta}{3}\Big(-\frac{2\Lambda_f}{D-6}+\frac{f^2}{2}\Big),
\eea
respectively.
Here $\zeta$ is defined as $\zeta:= \frac{\alpha^2}{2(D-6)}-\frac{D-6}{D-2}$.
In the case of $f=0$,
the solution in the scalar-tensor theory is reproduced
with the following replacements,
$\alpha=-(D-6)\beta$
and 
$\zeta=\frac{D-6}{2}c$.
The deficit or surplus angle of the internal space 
at $Y\to\infty$
is given by
\bea
\Delta \Omega^{(f)}_{D-5}
=\Omega_{D-5}
\Big[
1
-\Big\{
  \frac{D-6}{D-2}
\frac{1+\frac{D-6}{(D-2)\zeta}}{1+\frac{f^2}{6H^2}}
 \Big\}^{\frac{D-5}{2}}
\Big].
\eea
The induced metric thus turns out to be
\bea
ds_{\rm ind}^2
=-dt^2+c_0^2 e^{2Ht}\delta_{ij}dx^i dx^j
+\frac{D-6}{3H^2+\frac{f^2}{2}}
 d\Omega_{D-5}^2.
\eea
The junction conditions of the metric and the scalar field
give
\bea
\kappa_D^2\sigma_f=2H\sqrt{3\big(D-2+\frac{D-6}{\zeta}\big)}\,,~~~~~
\gamma_f=-\frac{D-6}{(D-2)\alpha}\,.
\eea
The form field strength is continuous across the brane
and hence 
we do not need to
introduce 
any coupling 
of it
with the brane matter.

\vspace{0.3cm}

For all the solutions discussed in this subsection,
there is a curvature singularity at $Y\rightarrow\infty$ 
where the Kretschmann invariant $R_{ABCD}R^{ABCD}$ diverges.
The singularity is due to the presence of a deficit (or surplus)
solid angle there \cite{mu}.
In the presence of the matter fields, we can eliminate
such a singularity for the particular coupling constant,
for example, for $c=c_\ast$ from Eq. (\ref{def_scalar})
in the case of the scalar-tensor theory.

\section{Spectrum of the tensor metric perturbations}

In this section, we discuss the tensor metric perturbations
with respect to the ordinary three-dimensional space
which 
satisfy the transverse-traceless (TT) conditions.
We focus on the existence of the zero mode
and the possible excitations of Kaluza-Klein (KK) modes.

\subsection{The five-dimensional model}

We briefly review the five-dimensional case.
The tensor perturbation for the metric Eq. (\ref{5d}) is given by
\bea
ds^2
=
A^2
\Big(-dt^2+c_0^2 e^{2Ht}\big(\delta_{ij}+h_{ij}\big)dx^i dx^j
+ dy^2\Big),
\eea
where $h_{ij}$ satisfies the TT conditions 
$h^{ij}{}_{,j}=h^i{}_i=0$.

Decomposing the tensor perturbations into the Fourier modes
$$h_{ij}=\int dm d^3 k f_m(y) \varphi_{m,k}(t)e^{ik_j x^j} {\hat e}_{ij},$$
where $\hat e_{ij}$ denotes two independent polarizations,
we find the equation of motion for each mode
\bea
&&f_m{}''+3\frac{A'}{A}
 f_m{}' =-m^2  f_m.
\eea
The solution of $\varphi_{m,k}$ is given by Eq. (\ref{4dmode}).
It is convenient to rewrite the bulk equation
into the form of the Schr\"{o}dinger equation
by introducing new variable $f_m=A^{-\frac{3}{2}}X_m$,
\bea
\Big[-\frac{d^2}{dy^2}
   +V(y)
\Big]X_m(y)
=m^2 X_m(y),
\eea
where
\bea
V(y)=
-3H 
\sqrt{1+\frac{|\Lambda_5|}{6H^2}}
\delta(y-y_b)
+\frac{9H^2}{4}
+\frac{360|\Lambda_5| H^4}
  {(24H^2e^{H|y|}-|\Lambda_5| e^{-H|y|})^2}.
\label{pot_5d}
\eea
The attractive delta function term
represents the contribution from the brane
and ensures the existence of the massless zero mode.
The zero and KK modes satisfy the  
normalization conditions
\bea
&&2\int_{y_b}^\infty dy 
A^3
f_0 f_0=1,\quad
2\int_{y_b}^\infty dy 
A^3
f_m f_{m'}=\delta(m-m'),
\eea
respectively,
where $m$, $m'\neq 0$ in the second relation.
The factor 2 in front of the integrals 
reflects the $Z_2$-symmetry with respect to the brane.

The mode function for the zero mode is given by $f_0=C$,
where 
\bea
C^{-2}~
&=&2\int_{y_b}^{\infty} dy
A^3
\nn
\\
&=&
\frac{\sqrt{6}}{|\Lambda_5|^{3/2}}
\Big\{
|\Lambda_5|^{1/2}
\sqrt{6H^2+|\Lambda_5|}
+6H^2\ln
\Big(
\frac{\sqrt{6H^2}}
 {\sqrt{|\Lambda_5|}+\sqrt{6H^2+|\Lambda_5|}}
\Big)
\Big\}.
\eea
The mode function gives $C^{2}\simeq \sqrt{\frac{|\Lambda_5|}{6}}$ 
at the low energy scales $H\ll |\Lambda_5|^{\frac{1}{2}}$,
while one finds $C^2\simeq \frac{3H}{2}$
at the high energy scales  $H\gg |\Lambda_5|^{\frac{1}{2}}$.
This was the well-known result obtained in Ref. \cite{langlois}.
The effective four-dimensional gravitational mass is given by
\bea
M_{4,\rm eff}^2&=& \frac{1}{\kappa_5^2 C^2}
\nn\\
&=&\frac{\sqrt{6}}{\kappa_5^2 |\Lambda_5|^{3/2}}
\Big\{
|\Lambda_5|^{1/2}
\sqrt{6H^2+|\Lambda_5|}
+6H^2\ln
\Big(
\frac{\sqrt{6H^2}}
 {\sqrt{|\Lambda_5|}+\sqrt{6H^2+|\Lambda_5|}}
\Big)
\Big\}.
\label{5eff}
\eea 
We find that there is a well-behaved low energy limit of 
Eq. (\ref{5eff}) with a fixed $|\Lambda_5|$, 
as $M_{\rm eff}^2= \frac{1}{\kappa_5^2 C^2} \Big|_{\frac{H} {|\Lambda_5|}\to 0}
=\frac{1}{\kappa_5^2}\sqrt{\frac{6}{|\Lambda_5|}}$.
Eq. (\ref{pot_5d}) shows that 
since the potential $V>\frac{9H^2}{4}$,
there is always the mass gap between the zero mode
and the continuum of KK modes given by $\frac{3}{2}H$.
Thus there is no excitation of light KK modes
of $0<m<\frac{3}{2}H$.

\subsection{The $D(>6)$-dimensional models}

In this subsection, we consider the tensor perturbations
in higher-dimensional warped compactifications
with the de Sitter universe.
The metric including tensor perturbations 
is given by
\bea
ds^2
=
A(y)^2
\Big[
-dt^2
+c_0^2 e^{2Ht}
\big(\delta_{ij}+h_{ij}\big)dx^i dx^j
+dY^2
+\omega_{ab}dz^adz^b
\Big],\label{kegan2}
\eea
where $\omega_{ab}$ denotes the metric of the
$(D-5)$-dimensional spherical dimensions
with a given radius in each case.

Decomposing the tensor perturbations into the Fourier modes
$$h_{ij}=\int dm d^3 k \sum_{\{L\}} f_m(Y)Y_{\{L\}}
\varphi_{m,k}(t)e^{ik_j x^j}  {\hat e}_{ij},$$
where the solution of $\varphi_{m,k}$ is given by Eq.~(\ref{4dmode}), 
and 
$m$, $\{L\}$
denote the KK mass and 
a set of quantum numbers
associated with the $(D-5)$ sphere,
respectively.
In particular, $L$ denotes the azimuthal quantum number, and 
$Y_{\{L\}}$ corresponds to the harmonic function on $S^{D-5}$,
which satisfies $\Delta_{D-5}Y_{\{L\}}=-L(L+D-6)Y_{\{L\}}$,
and is normalized as $\int d\Omega_{D-5}Y_{\{L\}}Y_{\{L'\}}
=\delta_{\{L\},\{L'\}}$. 

\vspace{0.2cm}

\underline{\it (1) In the case of the pure gravity}

\vspace{0.2cm}

We take the perturbation in the metric (\ref{nd}) so that the
perturbed line element is of the form (\ref{kegan2}), where
\bea
A(Y)=e^{-H\sqrt{\frac{3}{D-2}}|Y-Y_0|},\quad
\omega_{ab}dz^adz^b=\frac{D-6}{3H^2}d\Omega_{D-5}^2.
\eea
We find that each mode satisfies the equation of motion
\bea
&&\frac{H^2 }{A^{D-2} }
\frac{d}{dY}\Big(A^{D-2}
\frac{d}{dY}f_m
\Big)
=-\Big(m^2-\frac{3 L (L+D-6)H^2}{D-6}\Big) f_m.
\label{radnd}
\eea
The zero and KK modes satisfy the normalization conditions
\bea
&&2\int_{Y_0}^\infty dY\, \Omega_{D-5}
\Big(\frac{D-6}{3}\Big)^{\frac{D-5}{2}}
\frac{1}{H^{D-4}}
A^{D-2}
f_0 f_0=1,\quad
\nonumber\\
&&
2\int_{Y_0}^\infty dY\,\Omega_{D-5}
\Big(\frac{D-6}{3}\Big)^{\frac{D-5}{2}}
\frac{1}{H^{D-4}}
A^{D-2}
f_m f_m'=\delta(m-m'),\label{normnd}
\eea
where $m$, $m'\neq 0$ in the second relation.

By the redefinition of $f_m=H^{\frac{1}{2}}
A^{-\frac{D-2}{2}}
X_m$,
Eq. (\ref{radnd}) reduces to
\bea
\label{red_nd}
&&\Big[-\frac{d^2}{dY^2}
+
\frac{3(D-2)}{4}
H^2
-\sqrt{3} H \sqrt{D-2}\delta(Y-Y_0)
\Big]X_m
\nn\\
&=&
\Big(m^2-\frac{3H^2 L (L+D-6)}{D-2}\Big)
 X_m.
\eea
All modes of $m^2=\frac{3L(L+D-6)}{D-6}H^2$
give the normalizable solutions of Eq. (\ref{red_nd}).
Thus excitations of the massive
bound states associated with the angular dimensions
are allowed.
However, the masses of these modes are always greater than the critical mass
of the de Sitter spacetime.
This is summarized in Fig. 1.
On the other hand,
for $L=0$, 
the mass gap between the zero mode and the continuum of the KK modes
associated with the non-compact 
$y$-direction
is given by $\frac{\sqrt{3(D-2)}}{2}H$.
Thus it does not depend on the parameters of $M$ and $p$.
As the number of extra dimensions increases,
the mass gap also increases.
For the cases of $D=10$ and $11$,
we obtain the mass gaps of $\sqrt{6}H$ and $\frac{3\sqrt{3}}{2}H$,
respectively.
Thus although the brane involves the internal angular dimensions,
in the four-dimensional universe
no light massive mode is excited.
\begin{figure}[h]
\begin{center}
\includegraphics[keepaspectratio, scale=0.8]{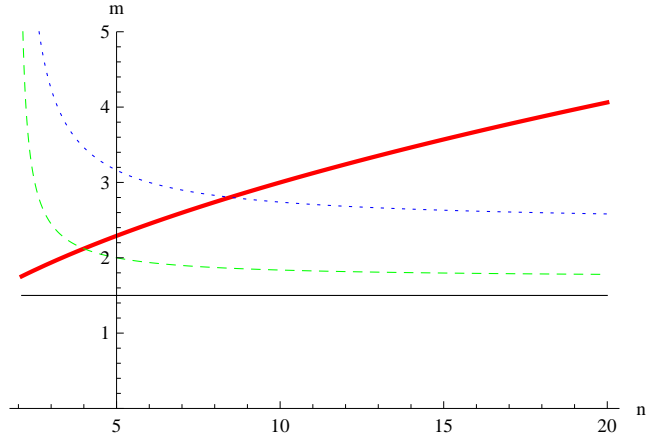}
\caption{The thick (red), dashed (green), dotted (blue)
curves represent the mass gap 
between the zero mode and
the continuum of
KK modes associated the $y$-direction,
and masses of the first and second massive bound states
associated with the angular directions, respectively,
in the unit of $H$
and 
as the functions of $n:=D-4$ which denotes
the dimensionality of the internal space.
The solid line shows the critical mass of the 
de Sitter spacetime.}
\label{fig1}
 \end{center}
\end{figure}
The zero mode solution is given by 
\bea
f_0= C_1+ C_2 A^{-(D-2)}.
\eea
The normalizability forbids the second solution
and we set $C_2=0$.
Then, through Eq. (\ref{normnd}), we obtain
\bea
\label{qqq}
C_1^2=\frac{\sqrt{3(D-2)}}{2\Omega_{D-5}}
\Big(\frac{3}{D-6}\Big)^{\frac{D-6}{2}}
H^{D-4}.
\eea
For $D=7$, we get $C_1^2=\frac{3\sqrt{15}H^3}{8\pi}$.
Also, 
in the limit of $H\to 0$,
the amplitude of the zero mode vanishes.
The effective four-dimensional gravitational mass is given by
\bea
M_{4,\rm eff}^2= \frac{1}{\kappa_D^2 C_1^2}
=\frac{2\Omega_{D-5}}{ \sqrt{3(D-2)} \kappa_D^2H^{D-4}}
\Big(\frac{D-6}{3}\Big)^{\frac{D-5}{2}},
\label{neff}
\eea 
which is ill-defined in the limit of $H\to 0$.

In order to see the difference from the analysis Ref. \cite{neupane2},
it is better to write down the bulk equation Eq. (\ref{radnd})
in terms of the original $y$ coordinate
\bea
&&\frac{H^2 
}{G^{\frac{1}{2}}A^{D-2} }
\frac{d}{dy}\Big(\frac{A^{D-2}}{
G^{\frac{1}{2}}}
\frac{d}{dy}f_m
\Big)
=-\Big(m^2-\frac{3H^2 L (L+D-6)}{D-6}\Big) f_m.
\eea
The equation \eqref{radnd} for $D=7$ and $L=0$
disagrees with Eq. (3.4) of Ref. \cite{neupane2}.
The difference should be due to the fact that 
in Eq. (3.4) of Ref. \cite{neupane2}, 
$m^2$ should be replaced with $m^2 G(y)$
in the separation of variables,
which is expected from Eq. (3.3).
Therefore,
the subsequent results and conclusions
are different.
For example,
in Ref. \cite{neupane2}
the potential (3.14) diverges at the brane position
and the final expression for the mass gap
becomes $\frac{25p^2M^2}{4}$
which explicitly depends on $p$ and $M$.
As we have mentioned in the previous section,
both $p$ and $M$ are not physical parameters.
This can be seen, for instance,
from the fact
that they do not appear in the metric after defining 
the new coordinate $Y$ as Eq. (\ref{nd}),
which 
is 
the proper coordinate in the conformal frame.
The physical coordinate $dZ=A(Y)dY$ also does not depend
on $p$ and $M$,
since $A(Y)$ does not depend on them.
In the $Y$ coordinate system (hence also in the $Z$ coordinate system),
except for the term of the delta function
representing the contribution from the brane,
the potential for the bulk eigen equation
is smooth at the brane.

In the case with a cosmological constant,
the metric including the tensor perturbations
is given by Eq. (\ref{kegan2}) with Eq. (\ref{cosmo}).
As in the previous case, 
decomposing into the Fourier modes 
we find that each mode satisfies Eq. (\ref{radnd}) with Eq. (\ref{ds_scale}).
By the redefinition of
$f_m=H^{\frac{1}{2}}
A^{-\frac{D-2}{2}}
X_m$,
the bulk mode function becomes
\bea
\label{red_nd_2}
&&\Big[-\frac{d^2}{dY^2}
+V(Y)
-\sqrt{D-2}\sqrt{3H^2+\frac{2|\Lambda_D|}{D-1}}\delta(Y-Y_0)
\Big]X_m
\nn\\
&=&\Big(m^2-\frac{3H^2 L (L+D-6)}{D-6}\Big)
 X_m.
\eea
The bulk potential $V(Y)$ is greater than
$\frac{3(D-2)H^2}{4}$,
monotonically decreases as increasing $Y$
and asymptotically approaches this value.
The behaviors of the modes are the same as
those
in the previous subsection,
which are summarized in Fig. 1.
All discrete modes of $m^2=\frac{3L(L+D-6)}{D-6}H^2$
are normalizable
and satisfy the boundary conditions. 
But for any $D$, all the massive bound states are heavier
than the critical mass of de Sitter $m_c$.  
On the other hand,
for $L=0$, 
the mass gap between the zero mode the continuum of KK modes
associated with the non-compact 
$y$-direction
is given by $\frac{\sqrt{3(D-2)}}{2}H$.
Thus the mass gap does not depend on the parameters of $M$ and $p$.
As the number of extra dimensions increases,
the mass gap also increases.

With
the normalizable zero mode solution
$f_0= C_1$,
the effective four-dimensional gravitational mass is given by
$M_{4,\rm eff}^2= \frac{1}{\kappa_D^2 C_1^2}$.
The normalization constant for $D=7$ is given by
\bea
C_1^{-2}
&=&\frac{\pi }{H^2|\Lambda_7|^{\frac{5}{2}}}
\sqrt{\frac{5}{3}}
\Big\{
\frac{243H^4}{2}
\ln\Big(\frac{\sqrt{|\Lambda_7|}+\sqrt{9H^2+|\Lambda_7|}}
         {-\sqrt{|\Lambda_7|}+\sqrt{9H^2+|\Lambda_7|}}\Big)
\nn\\
&-&(27H^2-2|\Lambda_7|)\sqrt{9H^2+|\Lambda_7|}|\Lambda_7|^{\frac{1}{2}}
\Big\}.
\eea
For $H\gg |\Lambda_7|^{\frac{1}{2}}$, 
the normalization constant becomes
$C_1^2\simeq \frac{3\sqrt{15}H^3}{8\pi}$,
while we find
$C_1^2\simeq \frac{H^2}{2\pi}\sqrt{\frac{3|\Lambda_7|}{5}}$, 
for $H\ll |\Lambda_7|^{\frac{1}{2}}$.
In the first case, we recovered the result in the previous
subsection.
Similarly for $D=11$, the normalization constant is
given by
\bea
C_1^{-2}
&=&\frac{25\sqrt{5}\pi^3}{216 H^6|\Lambda_{11}|^{\frac{9}{2}}}
\nonumber\\
&&\hspace{-1cm}\times
\Big\{
2|\Lambda_{11}|^{\frac{1}{2}}\sqrt{15H^2+|\Lambda_{11}|}
\big(-118125 H^6+5250 H^4|\Lambda_{11}|-280 H^2 |\Lambda_{11}|^2
+16|\Lambda_{11}|^3\big)
\nonumber\\
&+&1771875 H^8
\ln\Big(\frac{\sqrt{15H^2+|\Lambda_{11}|}+\sqrt{|\Lambda_{11}|}}
{\sqrt{15H^2+|\Lambda_{11}|}-\sqrt{|\Lambda_{11}|}}\Big)
\Big\}.
  \label{c1}
\eea
For $H\gg |\Lambda_{11}|^{\frac{1}{2}}$, Eq.~(\ref{c1}) gives
$C_1^2\simeq \frac{243\sqrt{3}H^7}{800\pi^3}$,
while we obtain $C_1^2\simeq \frac{27\sqrt{|\Lambda_{11}|} H^6}
{100\sqrt{5}\pi^3}$, for $H\ll |\Lambda_{11}|^{\frac{1}{2}}$.
The first limit coincides with Eq. (\ref{qqq}) for $D=11$.
In the $H\to 0$ limit, the zero mode amplitude vanishes.
Similar properties are obtained for any value of $D$.
The four-dimensional gravitational mass is still
ill-defined for $H\to 0$.

\vspace{0.2cm}

\underline{\it (2)  In the case of the scalar-tensor theory}

\vspace{0.2cm}

We then consider perturbations
about
the solution
in the scalar-tensor theory with the negative potential
\eqref{sol_scalar}.
The bulk mode obeys the eigen equation 
\bea
&&
\Big[-\frac{d^2}{dY^2}
+\Big(\frac{3\big(2+c(D-2)\big)H^2}{4c}
-H\sqrt{\frac{3\big(2+c(D-2)\big)}{c}}\delta(Y-Y_0)
\Big)
\Big]X_m
\nn\\
&&~~~=\Big(m^2-\frac{3H^2 L(L+D-6)}{D-6} \Big)X_m.
\eea
The solution for the zero mode is given by
\bea
X_0
\propto
e^{-\frac{1}{2}H
\sqrt{\frac{3(2+c(D-2))}{c}}|Y-Y_0|}.
\label{scalar_zero}
\eea
Other than the massless state, there are massive bound states
of $m^2=\frac{3H^2 L(L+D-6)}{D-6}$
which 
behave
as Eq. (\ref{scalar_zero}).
The effective four-dimensional gravitational mass is given by
\bea
M_{4,\rm eff}^2= \frac{1}{\kappa_D^2 C_1^2}
=\frac{2\Omega_{D-5}}{\kappa_D^2 H^{D-4}}
\Big(\frac{D-6}{3}\Big)^{\frac{D-5}{2}}
\sqrt{\frac{c}{3\big((D-2)c+2\big)}}.
\eea 
The mass gap between the zero mode and KK modes
associated with the warped dimensions is given by 
$\Delta m^2=\frac{3\big(2+c(D-2)\big)H^2}{4c}$,
which is bigger than the gap
in the case without a scalar field, 
$\frac{3(D-2)}{4}H^2$.
Thus in the case of the scalar-tensor theory,
it becomes more difficult
to excite the KK modes,
in particular for the limit $c\to 0$.

\vspace{0.2cm}

\underline{\it (3) In the case with a form field strength}

\vspace{0.2cm}

Finally, 
we consider perturbations 
about
the solution
in the theory including scalar and gauge fields
 Eq. (\ref{flux}).
The metric including perturbations 
is given by 
Eq. (\ref{kegan2})
with 
\bea
\omega_{ab}dz^adz^b=
\frac{D-6}{3H^2+\frac{f^2}{2}}d\Omega_{(D-5)}^2.
\eea
The bulk mode obeys the eigen equation
\bea
&&
\left[
-\frac{d^2}{dY^2}
+\frac{3(D-2)H^2}{4}\Big(1+\frac{D-6}{(D-2)\zeta}\Big)\right.\nn\\
&&\left.
~~-H\sqrt{3(D-2)\Big(1+\frac{D-6}{(D-2)\zeta}\Big)}\delta(Y-Y_0)
\right]X_m
\nonumber\\
&&~~~~~=\Big(m^2-\frac{3H^2 
\big(1+\frac{f^2}{6H^2}\big)
 L(L+D-6)}{D-6} \Big)X_m.
\eea
The normalizable solution for the zero mode is given by
\bea
X_0\propto
e^{-\frac{1}{2}H
\sqrt{3\big(D-2+\frac{D-6}{\zeta}\big)}|Y-Y_0|}.
\label{flux_zero}
\eea
Besides the zero mode, there are massive bound states
with $m^2=\frac{3H^2 L(L+D-6)}{D-6}\big(1+\frac{f^2}{6H^2}\big)$
which satisfy Eq. (\ref{flux_zero}).
The effective four-dimensional gravitational mass
is given by
\bea
M_{4,\rm eff}^2= \frac{1}{\kappa_D^2 C_1^2}
=\frac{2\Omega_{D-5}}{\kappa_D^2 H^{D-4}}
\Big(\frac{D-6}{3\big(1+\frac{f^2}{6H^2}\big)}\Big)^{\frac{D-5}{2}}
\frac{1}{\sqrt{3\big(D-2+\frac{D-6}{\zeta}\big)}}.
\eea 
There are two $H\to 0$ limits:
One is to set $\zeta=0$.
The other is to choose $f^2=\frac{4\Lambda_f}{D-6}$.
In the latter case,
the effective gravitational coupling diverges.
In the former case,
the effective gravitational
mass in the limit of $H\to 0$
is well-defined with
\bea
M_{4,\rm eff}^2=
\frac{2\Omega_{D-5}}{\kappa_D^2}
\Big(\frac{2(D-6)}{f^2}\Big)^{\frac{D-5}{2}}
\frac{1}{\sqrt{-2\Lambda_f+\frac{f^2}{2}(D-6)}}.
\label{mond}
\eea
The mass gap between the zero mode and KK modes
associated with the warped dimensions is given by 
\bea
\Delta m^2&=&\frac{3(D-2)H^2}{4}\big(1+\frac{D-6}{\zeta(D-2)}\big)
\nn\\
&=&\frac{1}{4}\Big(-\frac{2\Lambda_f}{D-6}+\frac{f^2}{2}\Big)
\big(\zeta(D-2)+D-6\big).
\eea
In the limit of $\zeta\to 0$,
the mass gap is still finite as
\bea
\Delta m^2=\frac{3(D-2)H^2}{4}\big(1+\frac{D-6}{\zeta(D-2)}\big)
=\frac{1}{4}\big(-2\Lambda_f+\frac{f^2(D-6)}{2}\big).
\eea 
In the case of $f^2=\frac{4\Lambda_f}{D-6}$,
there is no mass gap.
For a finite $H$, the mass gap is always greater
than $\frac{3(D-2)H^2}{4}$ in the case of the pure gravity.

\section{Scalar metric perturbations and stability}

Finally, we investigate stability 
against the scalar perturbations.
Here the term ``scalar'' is 
with respect to the four-dimensional de Sitter spacetime. 
It has been argued that 
the de Sitter solutions
become unstable against the scalar 
metric perturbations,
for example,
in the product spacetime of $dS_4\times S^n$ 
with flux \cite{kofman,krish}
and in the five-dimensional model with
two de Sitter branes \cite{gs,gs2}.
Thus, for these new de Sitter solutions \cite{neupane2,mu},
it is important to investigate the stability.

The most important purpose 
of
this section
is to see whether the lowest mode becomes tachyonic.
But then, it is also important to investigate
how large the negative mass square 
of 
a tachyonic mode is,
 even if 
it exists.
For an inflationary model,
the existence of a tachyonic mode whose 
absolute value of the negative mass square is much smaller than 
the Hubble parameter $H$
may be required to terminate inflation successfully
with an appropriate e-folding number of cosmic expansion.
Otherwise, inflation is terminated within a few Hubble time scales
and the model becomes unrealistic.
Thus, we will focus on the value of the lowest mode and 
also how its value is affected by the existence of the matter fields in the 
higher-dimensional theory.

Taking a longitudinal-type gauge,
the scalar perturbed metric is given by
\bea
ds^2= A(Y)^2\Big[
\big(1+2\phi_1\big)\gamma_{\mu\nu}dx^{\mu}dx^{\nu}
+\big(1+2\phi_2\big)dY^2
+\big(1+2\phi_3\big)\omega_{ab}dz^a dz^b
\Big],
\label{scalar_metric}
\eea
where $\gamma_{\mu\nu}$ is the de Sitter metric
and $\omega_{ab}$ is that for an $(D-5)$-sphere
of the given radius which is dependent on
the model.
$\phi_i$ ($i=1,2,3$) are the moduli of 
each direction.
The components of the Christoffel symbol and 
the curvature tensors obtained from the
metric Eq. (\ref{scalar_metric})
are summarized in the Appendix A.
We derive the equation which determines 
the stability, 
by combining the Einstein equations 
\bea
\delta G^A{}_B=\kappa_D^2 \delta T^A{}_B,
\eea
where $\delta T^A{}_B$ is 
perturbation of the energy-momentum tensor
of the bulk matter.

\subsection{In the case of the pure gravity}

Firstly, we consider the pure gravity solution
\eqref{nd0}.
Assuming the vacuum bulk and 
combining the Einstein
equations, we find
the equation
\bea
\big(\Box_4 +6H^2\big)\big(\phi_2-\phi_3\big)
-\sqrt{3(D-2)}H \big(\phi_2-\phi_3\big){}'
+\big(\phi_2-\phi_3\big){}''
=0.\label{master_eq}
\eea
The detailed derivation of this relation
is outlined in the Appendix B. 1.
This equation is the same as 
that for
the 
tensor perturbations,
except for the replacement of $\Box_4$
with $\big(\Box_4+6H^2\big)$.
Decomposing into the
effective four-dimensional modes,
$\phi_2-\phi_3=\int dm \psi_{m}(x^{\mu})g_m(Y)$,
the bulk eigen equation is the same as 
that
in the case of the 
tensor perturbations Eq. (\ref{radnd}).
Thus the zero mode of $m=0$ 
gives
$g_0(Y)=const$
and the effective four-dimensional equation
\bea
(\Box_4+6H^2)\psi_0=0,
\eea
and hence the lowest mode becomes tachyonic.
There is also the mass gap between the zero mode and higher modes
$\Delta m^2=\frac{3(D-2)}{4}H^2$ and the lowest KK mass is 
given by $m^2=\frac{3(D-10)}{4}H^2$.
For $D\geq 10$, all the KK modes are not tachyonic.
The absolute value of the tachyonic mass is too large
for the model \cite{neupane2} to be a realistic model.

We should also mention the normalizability of the scalar perturbations.
The normalization condition for the scalar perturbations
essentially remains the same as
that for the tensor modes:
$\int_{Y_0}^{\infty}dY A^{D-2}g_0(Y)^2< \infty $
for a discrete mode,
and also 
$\int_{Y_0}^{\infty}dY A^{D-2}g_{m}(Y)g_{m'}(Y)\propto 
\delta(m-m')$
for continuous modes $m\neq 0$ and $m'\neq 0$.
The boundary condition for $g_{m}(Y)$
on the brane which is obtained 
by integrating over it is 
also the same as that for the tensor perturbations.
Hence the normalizable solution for the 
tachyonic 
zero mode
in the scalar perturbations
is the same as that for the zero mode 
in the tensor perturbations,
and hence the corresponding mode
is also normalizable and physical.

\subsection{In the case of the scalar-tensor theory}

We then discuss the stability of the 
solution in the scalar-tensor theory
Eq. (\ref{sol_scalar}), outlined in the Appendix B. 2.
We 
obtain
the equation 
\bea
&&\Big(\Box_4+6H^2\Big)\big(\phi_2-\phi_3\big)
-\sqrt{3(D-2)\Big(1+\frac{2}{c(D-2)}\Big)}H
\big(\phi_2-\phi_3\big){}'
+\big(\phi_2{}-\phi_3{}\big)''
\nonumber\\
&&~~~~=0.
\label{master_eq2}
\eea
The differential operator has the same 
structure 
as that in the pure gravity model.
Decomposing into the effective four-dimensional modes,
the lowest mode becomes tachyonic with
the same mass square $-6H^2$.
Thus, this model is unstable against the scalar
perturbations.
Since the normalization and boundary conditions for a bulk mode
remain the same as those in the pure gravity model,
the tachyonic zero mode is also normalizable.
The lowest mass for the KK continuum
 is given by 
$\frac{3(2+(D-10)c)}{4c}H^2$.
The absolute value of the tachyonic mass is too large
for the model in the scalar-tensor theory
\cite{mu} to be a realistic model.

\subsection{In the case with a form field strength}

We finally consider
the solution
with a form field strength,
outlined in the Appendix B. 3.
For simplicity, we have ignored the 
perturbations of the form field strength.
The solution Eq.~(\ref{flux}) leads to the equation
\bea
&&\Big(\Box_4+6H^2\Big)\big(\phi_2-\phi_3\big)
-\sqrt{3(D-2)\Big(1+\frac{D-6}{\zeta(D-2)}\Big)}H
\big(\phi_2-\phi_3\big){}'
\nn
\\
&&~~~~+\big(\phi_2{}-\phi_3{}\big)''
=\frac{f^2}{2} \big(2\phi_3+\alpha\delta\phi\big).
\label{master3}
\eea
Although the differential operator in the left-hand side
is very similar to the previous cases,
the perturbation equation becomes
inhomogeneous due to the existence of the source term.
Such a source term is
induced for the nonvanishing flux.
In general the solution can be written 
in terms of the linear combination of 
the particular and homogeneous solutions.
From the homogeneous part,
the lowest mode becomes tachyonic with
the same mass square $-6H^2$.
Inclusion of the perturbations of the form field
modifies the form of the source term,
but does not affect the homogeneous,
hence geometrical
part of the equation Eq. (\ref{kum2}).
Thus, this model is also unstable against the scalar
perturbations.
Since the normalization and boundary conditions for a bulk mode
remain the same
as those in the previous two models,
the tachyonic zero mode is normalizable.
The lowest mass of the KK continuum is given by 
\bea
\frac{3}{4}H^2\Big(D-10+\frac{D-6}{\zeta}\Big).
\eea
The absolute value of the tachyonic mass is too large
for the model in the gravitational theory coupled to the form field
\cite{mu} to be a realistic model.

\subsection{
Interpretation of the tachyonic mode}

Although we find that
adding the matter degrees of freedom 
does not contribute to alleviate the instability,
our result suggests
that 
the value of the tachyonic mass in our model
relies on the geometrical (warped)
structure of the spacetime. 
If we can construct 
a more general class of solutions of 
the warped de Sitter compactification
where the spacetime structure
and the warp factor
can be determined
by several different mass scales,
it could provide a sufficiently large
parameter space
where a suppressed value of the tachyonic mass
is obtained,
even if a tachyonic mode exists.
Having such a class of solutions,
we could realize inflation with a sufficient e-folding number.

It should be mentioned that 
similar situations have been analyzed in the past works.
Let us review these models and then
clarify the difference of our model from them.
In Refs. \cite{gv1,gv2},
Garriga and Vilenkin analyzed
the fluctuations of a thin domain wall 
in the $(N+1)$-dimensional Minkowski spacetime. 
They showed that the wall 
fluctuation mode is represented 
by a scalar field living on the
$N$-dimensional de Sitter space 
which describes the internal metric on the domain wall, 
and the scalar field has
the negative mass squared $(-NH^2)$.
It turned out,
however,
that the wall fluctuation cannot be seen 
by an observer living on the wall 
because the fluctuation does not change 
the intrinsic curvature of the domain wall.
In addition, in Refs. \cite{gs,gs2}
it was shown
that in the system of a single de Sitter brane
in the five-dimensional spacetime
the fluctuations of the brane position 
cannot be seen by an observer on the brane,
while in the system of two de Sitter branes
the fluctuation of the relative displacement of branes
induces physically non-intrinsic effects on the brane,
since this is purely geometrical effect 
which can exist
without any source on the branes.
In other words,
only the relative displacement of branes can be
physical.

In contrast to the case of a domain wall 
and that of a single de Sitter brane in five dimensions,
as we have 
mentioned in the previous subsections,
in our model the scalar mode is normalizable and physical
even in the single brane system without any source on the brane.
Note that the important difference 
from the single-brane model in five dimensions 
is the 
existence of the internal $(D-5)$-sphere.
The scalar mode $\phi_2-\phi_3$
is clearly interpreted as the 
fluctuation of the size of the noncompact direction $Y$
relative to the $(D-5)$-sphere,
and this is in a situation similar to that of the relative 
displacement of two branes in five dimensions.
Thus an observer on the brane
would observe non-intrinsic effects from extra dimensions,
which must be revealed in the future studies.


Before closing this section,
we should also comment
on the moduli instability in
the lower-dimensional effective theory.
In Ref. \cite{mu},
we have analyzed
the lower-dimensional effective theory
obtained via integrating over
the spherical internal directions.
In the effective theory
two moduli fields appear,
which are associated with 
the size of the sphere
and the overall rescaling of the warp factor,
respectively.
It was found that
the first modulus
can be stabilized by the contribution of the field strength.
On the other hand,
the second modulus associated with the warp factor
cannot be stabilized 
by the classical ingredients in the original theory.
We expect 
that the instability of the scalar metric perturbations
found in this paper
corresponds to
the modulus instability in the lower-dimensional
 effective theory.
Thus the stabilization should be achieved by
some other mechanism.
A possible mechanism for
the stabilization 
is via the quantum corrections of the bulk matter fields,
which is under active study \cite{uzawa}.

\section{Conclusion}

In this paper, we have investigated the spectrum
of the tensor metric perturbations
from the warped solutions with 
the four-dimensional de Sitter spacetime
in the pure gravity,
in
the scalar-tensor theory
and 
also in
the theory with a form field strength.
The solutions 
have more than seven 
spacetime dimensions. 
In these solutions, the metric
is given by the warped product of 
the non-compact extra dimension, spherical extra dimensions
and the four-dimensional de Sitter spacetime.
To make the volume of extra dimensions finite, 
we construct a braneworld
using
the so-called cut-copy-paste method.
The difference from the five-dimensional case
is that 
the braneworld involves the spherical dimensions
as well as the de Sitter spacetime.
Thus
we 
compactify 
the spherical dimensions to obtain the 
four-dimensional cosmology.
The junction condition gives
the positive brane tension.

In all these models,
the tensor spectrum contains the massless zero mode.
In general, in the low energy limit
the amplitude of the zero mode vanishes,
which leads to the divergence of the four-dimensional
gravitational mass.
Adding a bulk cosmological constant does not
provide a resolution to this problem.
It implies that in such models
one 
cannot 
smoothly 
connect the de Sitter inflation
phase to the Friedmann universe.
Only the exceptional case 
is the case with a form field strength,
where the de Sitter expansion rate can be zero 
for a particular value of the coupling 
parameter to the field strength.

In the known five-dimensional case,
there is the mass gap between the zero and the continuum of KK modes,
which is equal to the critical mass of de Sitter spacetme. 
The mass gap between the zero mode and 
the continuum of the KK modes associated with the warped dimension
is greater than
the critical mass of the de Sitter spacetime.
In addition,
there exist
the massive bound states associated with 
the excitations along the spherical dimensions.
These modes are heavier than the critical mass in de Sitter spacetime.
Therefore,
although 
the braneworld involves the internal angular dimensions,
no light KK mode is excited.

We then have argued the stability of 
the solutions against the scalar perturbations
with respect to the four-dimensional symmetry.
We have shown 
the existence of the tachyonic
zero mode 
in all models.
The differential operator for the scalar perturbations
is the same as
that for
the tensor perturbations,
except that the four-dimensional operator is 
shifted by the tachyonic mass.
Irrespective of the presence of matter fields,
the mass of the lowest mode 
always takes $-6H^2$
which seems to be universal
and irrespective of the number of dimensions.

The absolute value of the tachyonic mass
is too large for 
the de Sitter compactifications 
discussed in Refs. \cite{neupane2,mu}
to be realistic inflationary models.
But if we can construct 
a more general class of solutions of 
the warped de Sitter compactification
where the spacetime structure and the warp factor
depend on several different mass scales,
we 
expect that
 it could give a sufficiently large parameter space
where
a suppressed value of the tachyonic mass
suitable for inflation with a sufficient e-folding number
is realized.
This subject is worth being investigated.
Finally,
we also mention that
the existence of the unstable mode is consistent with  
the analysis of the lower-dimensional effective theory. 
In the effective potential obtained after 
integrating over the spherical internal spaces,
the warp factor $A$ cannot be fixed,
which was originally found in Ref. \cite{mu}. 
We expect that
the instability of the scalar metric perturbations
corresponds
to such a modulus instability in the effective theory.
The stabilization via
the quantum effects of the bulk fields
is currently studied in \cite{uzawa}.
We hope to report these results in the future publication.

\section*{Acknowledgments}
K.U. was supported by Grant-in-Aid for 
Young Scientists (B) of JSPS Research, under Contract No. 20740147.
The work of M.M. was supported by Yukawa fellowship and 
by Grant-in-Aid for Young Scientists (B) of JSPS Research,
under Contract No. 24740162. 

 \vspace{1cm}

\appendix

\section{The components of the Christoffel symbol and the curvature tensors}

In this Appendix, we present the components of the 
Christoffel symbol and the curvature tensors obtained from 
the metric Eq. (\ref{scalar_metric})
which include the linear order scalar metric perturbations.

The components of the Christoffel symbol
obtained from the perturbed metric Eq. (\ref{scalar_metric})
are given by 
\bea
&&\Gamma^{Y}_{\mu\nu}
=-\frac{A'}{A}\gamma_{\mu\nu} 
 -\frac{A'}{A}\big(-2\phi_2+2\phi_1\big)\gamma_{\mu\nu}
 -\phi_1{}' \gamma_{\mu\nu},
\\
&&\Gamma^{Y}_{Y\mu}
=\phi_{2,\mu},
\\
&&\Gamma^Y_{YY}
=\frac{A'}{A}+\phi_2{}'
\\
&&\Gamma^{Y}_{ab}
=-\Big(
\frac{A'}{A}
+\frac{A'}{A}\big(2\phi_3-2\phi_2\big)
+\phi_3'
  \Big)\omega_{ab},
\\
&&
\Gamma^{Y}_{Ya}
=\phi_{2,a},
\\
&&
\Gamma^{\mu}_{YY}
=-\gamma^{\mu\nu}\phi_{2,\nu},
\\
&&
\Gamma^{a}_{YY}
=-\phi_{2,b}\omega^{ab},
\\
&&
\Gamma^{\mu}_{Y\nu}
=\delta^{\mu}_{\nu}
\Big(\frac{A'}{A}+\phi_1{}'\Big),
\\
&&
\Gamma^a_{Yb}
=\delta^{a}_{b}
\Big(\frac{A'}{A}+\phi_3{}'\Big),
\\
&&
\Gamma^{\mu}_{\alpha\beta}
={\tilde\Gamma}^{\mu}_{\alpha\beta}
+\big(
 \phi_{1,\beta}\delta^{\mu}_{\alpha}
+\phi_{1,\alpha}\delta^{\mu}_{\beta}
-\phi_{1,\nu}\gamma^{\mu\nu}\gamma_{\alpha\beta}
 \big),
\\
&&
\Gamma^{a}_{bc}
={\bar\Gamma}^{a}_{bc}
+\big(
 \phi_{1,b}\delta^{a}_{c}
+\phi_{1,c}\delta^{a}_{b}
-\phi_{1,d}\omega^{ad}\omega_{bc}
 \big),
\\
&&
\Gamma^{b}_{\mu\nu}
=-\gamma_{\mu\nu}\phi_{1,a}\omega^{ab},
\\
&&
\Gamma^{\alpha}_{ab}
=-\omega_{ab}\phi_{3,\beta}\gamma^{\beta\alpha},
\\
&&
\Gamma^{\mu}_{a\nu}
=\delta^{\mu}_{\nu}\phi_{1,a},
\\
&&
\Gamma^{a}_{\mu b}
=\delta^a_{b}\phi_{3,\mu},
\eea
where ${\tilde \Gamma}^{\mu}_{\alpha\beta}$
and ${\bar\Gamma}^{a}_{bc}$
are background Christoffel symbols computed from 
$\gamma_{\mu\nu}$ and $\omega_{ab}$,
respectively.

The components of the Riemann tensor are given by 
\bea
R^{Y}{}_{\mu Y \nu}
&=&-\Big(\frac{A'}{A}\Big)'\gamma_{\mu\nu}
-2\Big(\frac{A'}{A}\Big)'\big(-\phi_2+\phi_1\big)\gamma_{\mu\nu}
-\frac{A'}{A}\big(-\phi_2'+\phi_1'\big)\gamma_{\mu\nu}
\nn\\
&-&\phi_1{}'' \gamma_{\mu\nu}
-D_{\mu}D_{\nu}\phi_2,
\\
R^{Y}{}_{a Y b}
&=&-\Big(\frac{A'}{A}\Big)'\omega_{ab}
-2\Big(\frac{A'}{A}\Big)'\big(-\phi_2+\phi_3\big)\omega_{ab}
-\frac{A'}{A}\big(-\phi_2'+\phi_3'\big)\omega_{ab}
\nn\\
&-&\phi_3{}'' \omega_{ab}
-D_{a}D_{b}\phi_2,
\\
R^{\mu}{}_{Y\nu Y}
&=&-\Big(\frac{A'}{A}\Big)'\delta^{\mu}{}_{\nu}
-\delta^{\mu}{}_{\nu}\phi_1{}''
+\delta^{\mu}{}_{\nu}
 \frac{A'}{A}
 \big(\phi_2{}'-\phi_1{}'\big)\delta^{\mu}{}_{\nu}
 -D^{\mu}D_{\nu}\phi_2,
\\
R^{a}{}_{Yb Y}
&=&-\Big(\frac{A'}{A}\Big)'\delta^{a}{}_{b}
-\delta^{a}{}_{b}\phi_3{}''
+\delta^{a}{}_{b}
 \frac{A'}{A}
 \big(\phi_2{}'-\phi_3{}'\big)\delta^{a}{}_{b}
 -D^{a}D_{b}\phi_2,
\\
R^{\mu}{}_{\nu Y \alpha}
&=&\phi_{1,\nu}{}'\delta^{\mu}{}_{\alpha}
-\phi_{1,\rho}{}'\gamma^{\rho\mu}\gamma_{\nu\alpha}
+\frac{A'}{A}\phi_{2,\rho}\gamma^{\mu\rho}\gamma_{\nu\rho}
-\delta^{\mu}{}_{\alpha}\frac{A'}{A}\phi_{2,\nu},
\\
R^{a}{}_{b Y c}
&=&\phi_{3,b}{}'\delta^{a}{}_{c}
-\phi_{3,d}{}'\omega^{da}\omega_{bc}
+\frac{A'}{A}\phi_{2,d}\omega^{da}\omega_{bc}
-\delta^{a}{}_{c}\frac{A'}{A}\phi_{2,b},
\eea
\bea
R^{\alpha}{}_{\beta\mu\nu}
&=&{\tilde R}^{\alpha}{}_{\beta\mu\nu}
+\big(
 \delta^{\alpha}{}_{\mu}\gamma_{\beta\nu}
-\delta^{\alpha}{}_{\nu}\gamma_{\beta\mu}
 \big)
 \Big(\frac{A'}{A}\Big)^2
\nonumber\\
&+&D_{\mu}D_{\beta}\phi_1\delta^{\alpha}{}_{\nu}
-D_{\mu}D^{\alpha}\phi_1\gamma_{\nu\beta}
-D_{\nu}D_{\beta}\phi_1 \delta^{\alpha}{}_{\mu}
+D_{\nu}D^{\alpha}\phi_1 \gamma_{\mu\beta}
\nonumber\\
&-&
\big(
 \delta^{\alpha}{}_{\mu}\gamma_{\beta\nu}
-\delta^{\alpha}{}_{\nu}\gamma_{\beta\mu}
 \big)
\Big[
 \Big(\frac{A'}{A}\Big)^2(-2\phi_2+2\phi_1)
+2\phi_1'\frac{A'}{A}
\Big],
\\
R^{a}{}_{bcd}
&=&{\bar R}^{a}{}_{bcd}
+\big(
 \delta^{a}{}_{c}\omega_{bd}
-\delta^{a}{}_{d}\omega_{bc}
 \big)
 \Big(\frac{A'}{A}\Big)^2
\nonumber\\
&+&D_{c}D_{b}\phi_3\delta^{a}{}_{d}
-D_{c}D^{a}\phi_3\omega_{db}
-D_{d}D_{b}\phi_3 \delta^{a}{}_{c}
+D_{d}D^{a}\phi_3 \omega_{bc}
\nonumber\\
&-&
\big(
 \delta^{a}{}_{c}\omega_{bd}
-\delta^{a}{}_{d}\omega_{bc}
 \big)
\Big[
 \Big(\frac{A'}{A}\Big)^2(-2\phi_2+2\phi_3)
+2\phi_3'\frac{A'}{A}
\Big],
\\
R^{Y}{}_{\mu Y a}
&=&R^{Y}{}_{a Y\mu}
=-\phi_{2,\mu a},
\\
R^{a}{}_{\mu b\nu}
&=&-\Big(\frac{A'}{A}\Big)^2\delta^a{}_b\gamma_{\mu\nu}
\nonumber\\
&-&2\Big(\frac{A'}{A}\Big)^2 \big(\phi_1-\phi_2\big)
\delta^a{}_b\gamma_{\mu\nu}
-\frac{A'}{A}\big(\phi_1{}'+\phi_3{}'\big)\delta^{a}{}_{b}\gamma_{\mu\nu}
-\gamma_{\mu\nu}D^{a}D_{b}\phi_1
\nn\\
&-&\delta^{a}{}_{b}D_\mu D_\nu \phi_3,
\\
R^{\mu}{}_{a\nu b}
&=&-\Big(\frac{A'}{A}\Big)^2\delta^\mu{}_\nu\omega_{ab}
\nonumber\\
&-&2\Big(\frac{A'}{A}\Big)^2 \big(\phi_3-\phi_2\big)\delta^\mu{}_\nu\omega_{ab}
-\frac{A'}{A}\big(\phi_1{}'+\phi_3{}'\big)\delta^{\mu}{}_{\nu}\omega_{ab}
-\omega_{ab}D^{\mu}D_{\nu}\phi_3
\nn\\
&-&\delta^{\mu}{}_{\nu}D_a D_b \phi_1,
\\
R^{\alpha}{}_{\mu\beta a}
&=&-\phi_{1,\mu a}\delta^{\alpha}{}_{\beta}
+\phi_{1,\nu a}\gamma^{\nu \alpha}\gamma_{\mu\beta}
\\
R^{a}{}_{cb \mu}
&=&-\phi_{3,c \mu}\delta^{a}{}_{b}
+\phi_{3}{}^{,a}{}_{,\mu}\omega_{bc},
\\
R^{\alpha}{}_{a\beta\mu}
&=&\delta^{\alpha}{}_{\mu}\phi_{1,a\beta}
-\delta^{\alpha}{}_{\beta}\phi_{1,a\mu},
\\
R^a{}_{\mu bc}
&=&\delta^a{}_c\phi_{3,\mu b}
-\delta^a{}_b \phi_{3,\mu c},
\\
R^a{}_{Yb\mu}
&=&-\delta^a{}_b\phi_3{}'{}_{,\mu}
 +\delta^a{}_b \phi_{2,\mu}\frac{A'}{A},
\\
R^{\alpha}{}_{Y\beta a}
&=&-\delta^{\alpha}{}_{\beta}\phi_1{}'{}_{,a}
 +\delta^{\alpha}{}_\beta \phi_{2,a}\frac{A'}{A},
\\
R^{\alpha}{}_{Y\beta\gamma}
&=&
\delta^{\alpha}{}_{\gamma} \phi_{1,\beta}{}'
-\delta^{\alpha}{}_{\beta}\phi_{1,\gamma}{}'
+\delta^{\alpha}{}_{\beta}\frac{A'}{A}\phi_{2,\gamma}
-\delta^{\alpha}{}_{\gamma}\frac{A'}{A}\phi_{2,\beta},
\\
R^a{}_{Ybc}
&=&\delta^{a}{}_{c} \phi_{3,b}{}'
-\delta^{a}{}_{b}\phi_{3,c}{}'
+\delta^{a}{}_{b}\frac{A'}{A}\phi_{2,c}
-\delta^{a}{}_{c}\frac{A'}{A}\phi_{2,b},
\eea
where
${\tilde R}^{\alpha}{}_{\beta\mu\nu}$
and
${\bar R}^{a}{}_{bcd}$
are the background Riemann tensors with respect to
$\gamma_{\mu\nu}$
and 
$\omega_{ab}$,
respectively.

The components of the background Ricci tensor are given by
\bea
R^{(0)}_{YY}&=&-(D-1)\Big(\frac{A'}{A}\Big)',
\\
R^{(0)}_{\mu\nu}
&=&-\Big(\frac{A'}{A}\Big)'\gamma_{\mu\nu}
-(D-2)\Big(\frac{A'}{A}\Big)^2\gamma_{\mu\nu}
+{\tilde R}_{\mu\nu},
\\
R^{(0)}_{ab}
&=&-\Big(\frac{A'}{A}\Big)'\omega_{ab}
-(D-2)\Big(\frac{A'}{A}\Big)^2\omega_{ab}
+{\bar R}_{ab},
\eea 
where
${\tilde R}_{\mu\nu}$
and
${\bar R}_{ad}$
are the background Ricci tensors with respect to
$\gamma_{\mu\nu}$
and 
$\omega_{ab}$,
respectively.
Those of the perturbed Ricci tensor are given by
\bea
\delta R_{YY}&=&-\big(\Box_4+\Delta_{D-5}\big)\phi_2
  -4\phi_1{}''-(D-5)\phi_3{}''
+(D-1)\frac{A'}{A}\phi_2{}'
-4\frac{A'}{A}\phi_1{}'
\nn\\
&-&(D-5)\frac{A'}{A}\phi_3{}'\,,
\\
\delta R_{\mu\nu}
&=&-2D_{\mu}D_{\nu}\phi_1
-\gamma_{\mu\nu}\big(\Box_4+\Delta_{D-5}\big)\phi_1
-(D-5)D_{\mu}D_{\nu}\phi_3
-D_{\mu}D_{\nu}\phi_2
\nn\\
&-&2(D-2)\Big(\frac{A'}{A}\Big)^2(\phi_1-\phi_2)\gamma_{\mu\nu}
-(D+2)\frac{A'}{A}\phi_1{}'\gamma_{\mu\nu}
-(D-5)\frac{A'}{A}\phi_3{}' \gamma_{\mu\nu}
\nn\\
&+&\frac{A'}{A}\phi_2{}' \gamma_{\mu\nu}
-\phi_1{}'' \gamma_{\mu\nu}
-2\Big(\frac{A'}{A}\Big)'
\big(-\phi_2+\phi_1\big)\gamma_{\mu\nu}\,,
\\
\delta R_{ab}
&=&-2D_{a}D_{b}\phi_3
-\omega_{ab}\big(\Box_4+\Delta_{D-5}\big)\phi_3
-4D_{a}D_{b}\phi_1
-D_{a}D_{b}\phi_2
\nn \\
&-&
2(D-2)\Big(\frac{A'}{A}\Big)^2(\phi_3-\phi_2)\omega_{ab}
-(2D-7)\frac{A'}{A}\phi_3{}'\omega_{ab}
-4\frac{A'}{A}\phi_1{}' \omega_{ab}
\nn\\
&+&\frac{A'}{A}\phi_2{}' \omega_{ab}
-\phi_3{}'' \omega_{ab}
-2\Big(\frac{A'}{A}\Big)'
\big(-\phi_2+\phi_3\big)\omega_{ab},
\\
\delta R_{\mu a}
&=&-\Big(\phi_2+3\phi_1+(D-6)\phi_3\Big)_{,\mu a}\,,
\\
\delta R_{Y\mu}
&=&-3\phi_{1,\mu}{}'
+(D-2)\frac{A'}{A}\phi_{2,\mu}
-(D-5)\phi_{3,\mu}{}'\,,
\\
\delta R_{Y a}
&=&
-(D-6)\phi_{3,a}{}'
+(D-2)\frac{A'}{A}\phi_{2,a}
-4\phi_{1,a}{}'.
\eea

\section{The derivation of the equations of motion
for the scalar perturbations}

In this Appendix, we briefly summarize the equations of motion 
for the scalar perturbations. 
In what follows, we use the results shown in the Appendix A.

\subsection{In the case of the pure gravity}

In the case of the pure gravity model,
assuming the metric form Eq. (\ref{scalar_metric}),
it is straightforward to confirm that 
\bea
&&
A=e^{-\sqrt{\frac{3}{D-2}}H |Y-Y_0|},\quad
\omega_{ab}dz^a dz^b
=\frac{D-6}{3H^2} d\Omega_{(D-5)}^2,\quad
\nn\\
&&\gamma_{\mu\nu}dx^{\mu}dx^{\nu}
=-dt^2+c_0^2 e^{2Ht}\delta_{ij}dx^i dx^j,
\eea
is the solution to the background equations $R^{(0)}_{AB}=0$,
where we have employed 
${\tilde R}_{\mu\nu}=3H^2 \gamma_{\mu\nu}$
and ${\bar R}_{ab}=3H^2 \omega_{ab}$.

Taking the property of the background equation
$\frac{A'}{A}=-\sqrt{\frac{D-2}{3}}H$
and 
$\big(\frac{A'}{A}\big)'=0$,
the components of the perturbed Einstein tensor 
associated with the metric (\ref{scalar_metric})
are
given by
\bea
A^2\delta G^{\mu}{}_{\nu}
&=&-D^{\mu}D_{\nu}
\big(2\phi_1+(D-5)\phi_3+\phi_2\big)
+\Big[
\Box_{4}
\big(
2\phi_1+(D-5)\phi_3+\phi_2
\big)
\nn\\
&+&\Delta_{D-5}
\big(
3\phi_1+\phi_2+(D-6)\phi_3
\big)
\nonumber\\
&+&
6H^2
\big(\phi_1-\phi_2\big)
+3(D-5)H^2
\big(\phi_3-\phi_2\big)
\nn\\
&-&\sqrt{3(D-2)}H
\Big(3\phi_1{}'
    +(D-5)\phi_3{}'
    -\phi_2{}'
\Big)
+3\phi_1{}''+(D-5)\phi_3{}''
\Big]\delta^\mu{}_\nu,
 \\
A^2\delta G^{a}{}_{b}
&=&-D^{a}D_{b}
\big(4\phi_1+(D-7)\phi_3+\phi_2\big)
+\Big[
\Delta_{D-5}
\big(
4\phi_1+(D-7)\phi_3+\phi_2
\big)
\nonumber\\
&+&\Box_4
\big(
3\phi_1+\phi_2+(D-6)\phi_3
\big)
\nonumber\\
&+&
12 H^2
\big(\phi_1-\phi_2\big)
+
3(D-7)H^2
\big(\phi_3-\phi_2\big)
\nn\\
&-&\sqrt{3(D-2)}H
\Big(4\phi_1{}'
    +(D-6)\phi_3{}'
    -\phi_2{}'
\Big)
+4\phi_1{}''+(D-6)\phi_3{}''
\Big]\delta^a{}_b,
\\
A^2\delta G^{Y}{}_{Y}
&=&\Box_4 \Big( 3\phi_1+ (D-5)\phi_3\Big)
+\Delta_{D-5}\Big(4\phi_1+(D-6)\phi_3\Big)
\nonumber\\
&-&\sqrt{3(D-2)}H
\Big(4\phi_1{}'+(D-5)\phi_3{}'\Big)
\nn\\
&+&
3H^2
\Big(4\phi_1+(D-5)\phi_3-(D-1)\phi_2\Big),
\nonumber\\
\delta G_{\mu a}
&=&-\Big(\phi_2+3\phi_1+(D-6)\phi_3\Big)_{,\mu a},
\\
\delta G_{Y\mu}
&=&-3\phi_{1,\mu}{}'
-\sqrt{3(D-2)}H
\phi_{2,\mu}
-(D-5)\phi_{3,\mu}{}'\,,
\\
\delta G_{Y a}
&=&
-(D-6)\phi_{3,a}{}'
-\sqrt{3(D-2)}H
\phi_{2,a}
-4\phi_{1,a}{}'\,.
\eea

The components of the perturbed Einstein equations are given by
\bea
0&=&-D^{\mu}D_{\nu}
\big(2\phi_1+(D-5)\phi_3+\phi_2\big)
\nn
\\
&+&\Big[
\Box_{4}
\big(
2\phi_1+(D-5)\phi_3+\phi_2
\big)
+\Delta_{D-5}
\big(
3\phi_1+\phi_2+(D-6)\phi_3
\big)
\nonumber\\
&+&
6H^2
\big(\phi_1-\phi_2\big)
+ 3(D-5)H^2
\big(\phi_3-\phi_2\big)
\nn\\
&-&
H\sqrt{3(D-2)}
\Big(3\phi_1{}'
    +(D-5)\phi_3{}'
    -\phi_2{}'
\Big)
+3\phi_1{}''+(D-5)\phi_3{}''
\Big]\delta^\mu{}_\nu,
 \\
0&=&-D^{a}D_{b}
\big(4\phi_1+(D-7)\phi_3+\phi_2\big)
\nn\\
&+&\Big[
\Delta_{D-5}
\big(
4\phi_1+(D-7)\phi_3+\phi_2
\big)
+\Box_4
\big(
3\phi_1+\phi_2+(D-6)\phi_3
\big)
\nonumber\\
&+&
12H^2 
\big(\phi_1-\phi_2\big)
+ 3(D-7)H^2
\big(\phi_3-\phi_2\big)
\nn\\
&-&\sqrt{3(D-2)}H
\Big(4\phi_1{}'
    +(D-6)\phi_3{}'
    -\phi_2{}'
\Big)
+4\phi_1{}''+(D-6)\phi_3{}''
\Big]\delta^a{}_b,
\\
0&=&\Box_4 \Big( 3\phi_1+ (D-5)\phi_3\Big)
+\Delta_{D-5}\Big(4\phi_1+(D-6)\phi_3\Big)
\nn\\
&-&H\sqrt{3(D-2)}
\Big(4\phi_1{}'+(D-5)\phi_3{}'\Big)
\nn\\
&+&
3H^2 
\Big(4\phi_1+(D-5)\phi_3-(D-1)\phi_2\Big),
\eea
\bea
0&=&-\Big(\phi_2+3\phi_1+(D-6)\phi_3\Big)_{,\mu a},
\\
0&=&-3\phi_{1,\mu}{}'
-\sqrt{3(D-2)}H
\phi_{2,\mu}
-(D-5)\phi_{3,\mu}{}'\,,
\\
0&=&
-(D-6)\phi_{3,a}{}'
-\sqrt{3(D-2)}H
\phi_{2,a}
-4\phi_{1,a}{}'.
\eea
Here, we ignore the massive excitations in the $S^{D-5}$ direction
and set  
the terms with $\partial_a$ to be zero.
Then, the perturbed Einstein equations reduce simply to
\bea
\label{munu}
0&=&-D^{\mu}D_{\nu}
\big(2\phi_1+(D-5)\phi_3+\phi_2\big)
+\Big[
\Box_{4}
\big(
2\phi_1+(D-5)\phi_3+\phi_2
\big)
\nonumber\\
&+&
6H^2
\big(\phi_1-\phi_2\big)
+3(D-5)H^2
\big(\phi_3-\phi_2\big)
\nn
\\
&-&H\sqrt{3(D-2)}
\Big(3\phi_1{}'
    +(D-5)\phi_3{}'
    -\phi_2{}'
\Big)
+3\phi_1{}''+(D-5)\phi_3{}''
\Big]\delta^\mu{}_\nu,
 \\
\label{ab}
0&=&
\Big[
\Box_4
\big(
3\phi_1+\phi_2+(D-6)\phi_3
\big)
\nonumber\\
&+&
12H^2 
\big(\phi_1-\phi_2\big)
+ 3(D-7)H^2
\big(\phi_3-\phi_2\big)
\nn\\
&-&\sqrt{3(D-2)}H
\Big(4\phi_1{}'
    +(D-6)\phi_3{}'
    -\phi_2{}'
\Big)
+4\phi_1{}''+(D-6)\phi_3{}''
\Big]\delta^a{}_b,
\\
\label{yy}
0&=&\Box_4 \Big( 3\phi_1+ (D-5)\phi_3\Big)
-H\sqrt{3(D-2)}
\Big(4\phi_1{}'+(D-5)\phi_3{}'\Big)
\nonumber\\
&+&3H^2 
\Big(4\phi_1+(D-5)\phi_3-(D-1)\phi_2\Big),
\\
\label{ymu}
0&=&-3\phi_{1,\mu}{}'
-\sqrt{3(D-2)}H
\phi_{2,\mu}
-(D-5)\phi_{3,\mu}{}',
\eea
where $(\mu,a)$ and $(Y,a)$ components become trivial.

We assume that there is no anisotropic stress in the four dimensions
and $\delta T_{Y\mu}=0$. 
Eq.~(\ref{ymu}) gives
\bea
3\phi_1{}'+(D-5)\phi_3{}'
+\sqrt{3(D-2)}H\phi_2=0.\label{gogo}
\eea
We choose $\phi_1$ so that $\phi_1$ obeys the equation
\bea
2\phi_1=-(D-5)\phi_3-\phi_2.\label{zero_aniso}
\eea
Combined Eq. (\ref{gogo})
with Eq. (\ref{zero_aniso}), we obtain
\bea
3\phi_2{}'+(D-5)\phi_3{}'
-2\sqrt{3(D-2)}H\phi_2=0.
\label{come}
\eea
Taking the difference between the trace of Eq. (\ref{ab})
and Eq. (\ref{yy}), we get
\bea
0
&=&\Box_4 \big(\phi_2-\phi_3\big)
+6H^2 (\phi_2-\phi_3)
+\sqrt{3(D-2)}H \big(\phi_2{}'+\phi_3{}'\big)
+4\phi_1{}''+(D-6)\phi_3{}''
\nonumber\\
&=&
\Box_4 \big(\phi_2-\phi_3\big)
+6H^2 (\phi_2-\phi_3)
+\sqrt{3(D-2)}H \big(\phi_2{}'+\phi_3{}'\big)
\nn\\
&-&\Big(2\phi_2{}''+(D-4)\phi_3{}''\Big),
\label{heng}
\eea
where we have used Eq. (\ref{zero_aniso}).
Furthermore, we find
\bea
2\phi_2{}''+(D-4)\phi_3{}''
&=&
-(\phi_2{}''-\phi_3{}'')
+3\phi_2{}''+(D-5)\phi_3{}''
\nonumber\\
&=&
-(\phi_2-\phi_3)''
+2\sqrt{3(D-2)}H\phi_2{}',
\eea
where we have used Eq. (\ref{come}).
Combining with Eq. (\ref{heng}), we
find
\bea
0
&=&
\big(\Box_4 +6H^2\big)\big(\phi_2-\phi_3\big)
-\sqrt{3(D-2)}H \big(\phi_2-\phi_3\big){}'
\nonumber\\
&+&\big(\phi_2-\phi_3\big){}''.
\eea

\subsection{In the case of the scalar-tensor theory}

We then consider the perturbations 
about
the solution
in the scalar-tensor theory Eq. (\ref{sol_scalar}).
We consider the perturbations of the scalar field 
as well as the metric.
For the metric form (\ref{scalar_metric}),
the background solution of Einstein equations
is obtained as
\bea
&&A=e^{-\sqrt{\frac{3\big(\frac{2}{D-2}+c\big)}{c(D-2)}}H |Y-Y_0|},\quad
\omega_{ab}dz^a dz^b
=\frac{D-6}{3H^2} d\Omega_{(D-5)}^2,\quad
\nonumber\\
&&\gamma_{\mu\nu}dx^{\mu}dx^{\nu}
=-dt^2+c_0^2 e^{2Ht}\delta_{ij}dx^i dx^j\,.
\eea
Also,
${\tilde R}_{\mu\nu}=3H^2 \gamma_{\mu\nu}$
and ${\bar R}_{ab}=3H^2 \omega_{ab}$.

The energy-momentum tensor is given by
\bea
\kappa_D^2T_{AB}
=\frac{1}{2}\partial_A\phi\partial_B\phi
-\frac{1}{2}g_{AB}
\Big(
\frac{1}{2}g^{CD}\partial_C \phi
                 \partial_D \phi
-2e^{\beta\phi}
\Lambda_s
\Big).
\eea
The components of the background 
energy-momentum tensor 
are given by
\bea
\kappa_D^2 T^{(0)Y}{}_Y
&=&\frac{1}{4A^2}\phi'{}^2
+e^{\beta\phi}\Lambda_s,
\nonumber\\
\kappa_D^2 T^{(0)\mu}{}_\nu
&=&
\Big[-\frac{1}{4A^2}\phi'{}^2 
+e^{\beta\phi}\Lambda_s
\Big]
\delta^\mu{}_\nu,
\nonumber\\
\kappa_D^2 T^{(0)a}{}_b
&=&
\Big[-\frac{1}{4A^2} \phi'{}^2
+e^{\beta\phi}\Lambda_s
\Big]
\delta^a{}_b.
\eea
The components of the perturbed energy-momentum tensor 
are given by 
\bea
\kappa_D^2 \delta T^Y{}_Y
&=&\frac{1}{2A^2}\big(\phi'\delta\phi'-\phi_2 \phi'{}^2\big)
+\beta e^{\beta\phi}\Lambda_s 
\delta\phi,
\nonumber\\
\kappa_D^2 \delta T^\mu{}_\nu
&=&
\Big[-\frac{1}{2A^2}\big(\phi'\delta\phi'-\phi_2 \phi'{}^2\big)
+\beta e^{\beta\phi}\Lambda_s
\delta\phi\Big]
\delta^\mu{}_\nu,
\nonumber\\
\kappa_D^2 \delta T^a{}_b
&=&
\Big[-\frac{1}{2A^2}\big(\phi'\delta\phi'-\phi_2 \phi'{}^2\big)
+\beta e^{\beta\phi}\Lambda_s
\delta\phi\Big]
\delta^a{}_b,
\nonumber \\
\kappa_D^2 
\delta T_{Y\mu}&=&
\frac{1}{2}\phi'\delta \phi_{,\mu},
\eea
where we have ignored the excitations along $S^{D-5}$
directions and 
set the terms with $\partial_a$ to be zero.
Hence,
\bea
\kappa_D^2
\delta \Big(\frac{1}{D-5}T^{a}{}_a-T^Y{}_Y\Big)
=\frac{1}{A^2}\Big(\phi_2\phi'{}^2-\phi'\delta\phi'\Big).
\label{diff_st}
\eea
Now we derive the equation for the perturbations.
We consider the following combination
\bea
\delta \Big(\frac{1}{D-5}G^{a}{}_a-G^Y{}_Y\Big)
&=&\frac{1}{D-5}g^{(0)ac}\delta R_{ac}
-g^{YY}\delta R_{YY}
+\frac{1}{D-5}\delta g^{ac} R^{(0)}_{ac}
\nonumber\\
&&\hspace{-1.5cm}=A^{-2}
\Big[
\Box_4\big(\phi_2-\phi_3\big)
-(D-2)\frac{A'}{A}\big(\phi_2{}'+\phi_3{}'\big)
+4\phi_1{}''
+(D-6)\phi_3{}''
\nonumber\\
&&\hspace{-1cm}+2(D-2)\Big(\frac{A'}{A}\Big)^2
 \big(\phi_2-\phi_3\big)
\Big]
-2\phi_3\frac{R^{(0)a}{}_a}{D-5}
\nonumber\\
&&\hspace{-1.5cm}
=A^{-2}
\Big[
\Box_4\big(\phi_2-\phi_3\big)
+(D-2)\frac{A'}{A}\big(\phi_2{}'-\phi_3{}'\big)
+\phi_2{}''
-\phi_3{}''
\nonumber\\
&&\hspace{-1cm}+2(D-2)\Big(\frac{A'}{A}\Big)^2
 \big(\phi_2-\phi_3\big)
-\phi'\delta\phi'
\Big]
-2\phi_3\frac{R^{(0)a}{}_a}{D-5},\label{diff_eins}
\eea
where we have used the $(T,\mu)$ and 
the off-diagonal part of $(\mu,\nu)$ components
of the perturbed Einstein equation
\bea
&&\frac{1}{2}\phi'\delta\phi_{,\mu}
=\Big(-3\phi_1{}'
 +(D-2)\frac{A'}{A}\phi_2
 -(D-5)\phi_3\Big)_{,\mu},
\nonumber\\
&&2\phi_1+(D-5)\phi_3+\phi_2=0.
\eea
For the case of the solution Eq. (\ref{sol_scalar}), we obtain
\bea
2\phi_3\frac{R^{(0)a}{}_a}{D-5}+A^{-2}\phi'{}^2\phi_2
=\frac{12H^2}{c(D-2)}\frac{1}{A^2}\big(\phi_2-\phi_3\big).
\label{kol}
\eea
From 
Eqs. (\ref{diff_st}) and (\ref{diff_eins}) 
with Eq. (\ref{kol}),
we find
\bea
&&\Big(\Box_4+6H^2\Big)\big(\phi_2-\phi_3\big)
-\sqrt{3(D-2)\Big(1+\frac{2}{c(D-2)}\Big)}H
\big(\phi_2-\phi_3\big){}'
\nonumber\\
&+&\big(\phi_2{}-\phi_3{}\big){}''
=0.
\label{kum}
\eea

\subsection{In the case with a form field strength}

We finally consider the perturbations of the theory including the form field strength Eq. (\ref{fins}).
Here, we simply consider the perturbations of the scalar field $\phi\to \phi+
\delta\phi$ as well as the metric perturbations.
For the metric form (\ref{scalar_metric}), 
the background solution of the Einstein equations is obtained as
\bea
&&A=e^{-\sqrt{\frac{3}{D-2}\big(1+\frac{D-6}{\zeta(D-2)}\big)}H |Y-Y_0|},\quad
\omega_{ab}dz^a dz^b
=\frac{D-6}{3H^2+\frac{f^2}{2}} d\Omega_{(D-5)}^2,\quad
\nonumber\\
&&\gamma_{\mu\nu}dx^{\mu}dx^{\nu}
=-dt^2+c_0^2 e^{2Ht}\delta_{ij}dx^i dx^j\,.
\eea
Also,
${\tilde R}_{\mu\nu}=3H^2 \gamma_{\mu\nu}$
and ${\bar R}_{ab}=\big(3H^2+\frac{f^2}{2}\big) \omega_{ab}$.
For simplicity, we set the perturbations of the form field
strength to be zero, but expect that this would not 
change the structure of the evolution equation for the radionic mode.
The energy-momentum tensor obtained from Eq. (\ref{fins}) 
is given by 
\bea
\kappa_D^2T_{AB}
&=&
-e^{-\frac{\alpha\phi}{D-6}}\Lambda_f g_{AB}
+
\frac{1}{2}\partial_A\phi\partial_B\phi
-\frac{1}{4}g_{AB}
g^{CD}\partial_C \phi
                 \partial_D \phi
\nonumber\\
&+&\frac{1}{2(D-5)!}e^{\alpha\phi}
\Big[
(D-5)F_{AM_1\cdots M_{D-5}}
F_{B}{}^{M_1\cdots M_{D-5}}
\nn\\
&-&\frac{1}{2}g_{AB}
F_{M_1\cdots M_{(D-5)}}F^{M_1\cdots M_{D-5}}
\Big].
\eea
The background components of the energy-momentum tensor 
are given by 
\bea
\kappa_D^2 A^2 T^{(0)Y}{}_Y
&=&
\frac{3H^2}{2}(D-1)\Big(\frac{D-6}{\zeta(D-2)}\Big)
-\frac{D-5}{4}f^2,
\nonumber\\
\kappa_D^2 A^2 T^{(0)a}{}_b
&=&
\Big[
\frac{3H^2}{2}(D-3)\Big(\frac{D-6}{\zeta(D-2)}\Big)
-\frac{D-7}{4}f^2\Big]\delta^a{}_b,
\nonumber \\
\kappa_D^2 A^2 T^{(0)\mu}{}_\nu
&=&
\Big[
\frac{3H^2}{2}(D-3)\Big(\frac{D-6}{\zeta(D-2)}\Big)
-\frac{D-5}{4}f^2\Big]\delta^\mu{}_\nu,
\eea
where 
we have used
\bea
&&e^{\alpha\phi}F_{a_1\cdots a_{D-5}}F^{a_1\cdots a_{D-5}}
=\frac{(D-5)!f^2}{A^2},\quad
\frac{\phi'{}^2}{4}
=\frac{3H^2}{2\zeta}\frac{D-6}{D-2},
\nonumber\\
&&\Lambda_f=-\frac{3(D-6)H^2}{2\zeta}
+\frac{(D-6)f^2}{4}.
\label{lambdaf}
\eea
Ignoring the excitations along the $S^{D-5}$ direction,
the components of the perturbed energy-momentum tensor
reduce to
\bea
\kappa_D^2A^2\delta T^y{}_y
&=&\frac{1}{2}\big(\phi'\delta\phi'-\phi_2 \phi'{}^2\big)
-\frac{f^2}{4}\alpha\delta\phi
+\frac{\alpha\Lambda_f}{D-2}\delta\phi
,
\nonumber\\
\kappa_D^2 A^2 \delta T^{a}{}_b
&=&\Big[-\frac{1}{2}\big(\phi'\delta\phi'-\phi_2 \phi'{}^2\big)
+\frac{f^2}{4}\alpha\delta\phi
+\frac{\alpha\Lambda_f}{D-2}\delta\phi
\Big]\delta^a{}_b,
\nonumber\\
\kappa_D^2 A^2\delta T^{\mu}{}_\nu
&=&
\Big[-\frac{1}{2}\big(\phi'\delta\phi'-\phi_2 \phi'{}^2\big)
-\frac{f^2}{4}\alpha\delta\phi
+\frac{\alpha\Lambda_f}{D-2}\delta\phi
\Big]\delta^\mu{}_\nu,
\nonumber\\
\kappa_D^2\delta T_{Y\mu}
&=&\frac{1}{2}\phi'\delta\phi_{,\mu},
  \label{pem}
\eea
where others are zero and 
we have ignored the perturbations of the form field.
Hence, Eq.~(\ref{pem}) gives 
\bea
\kappa_D^2\delta \Big(\frac{1}{D-5}T^{a}{}_a-T^Y{}_Y\Big)
=\frac{1}{A^2}\Big(\phi_2\phi'{}^2-\phi'\delta\phi'\Big)
+\frac{f^2}{2}\frac{\alpha}{A^2}\delta\phi.
\label{diff_st2}
\eea
We now derive the evolution equation for the perturbations.
As for the previous case,
we compute the following combination
\bea
\delta \Big(\frac{1}{D-5}G^{a}{}_a-G^Y{}_Y\Big)
&=&
A^{-2}
\Big[
\Box_4\big(\phi_2-\phi_3\big)
+(D-2)\frac{A'}{A}\big(\phi_2{}'-\phi_3{}'\big)
+\phi_2{}''
-\phi_3{}''
\nonumber\\
&+&2(D-2)\Big(\frac{A'}{A}\Big)^2
 \big(\phi_2-\phi_3\big)
-\phi'\delta\phi'
\Big]
-2\phi_3\frac{R^{(0)a}{}_a}{D-5},\label{diff_eins2}
\eea
where we have used the $(T,\mu)$ and 
the off-diagonal part of $(\mu,\nu)$ components
of the perturbed Einstein equation.
For the case of the solution Eq. (\ref{flux}), we can write
\bea
2\phi_3\frac{A^2 R^{(0)a}{}_a}{D-5}+\phi'{}^2\phi_2
&=&
-\Big[
\frac{6(D-6)}{(D-2)\zeta}H^2-f^2
\Big]\phi_3
+\frac{6(D-6)}{(D-2)\zeta}H^2\phi_2
\nonumber\\
&=&f^2\phi_3
+\frac{6(D-6)H^2}{(D-2)\zeta}
\big(\phi_2-\phi_3\big),
\label{kol2}
\eea
where we have used
\bea
A^2 R^{(0)a}{}_a=
-\frac{3(D-6)}{(D-2)\zeta}H^2+\frac{f^2}{2}.
\eea
From 
Eqs. (\ref{diff_st2}) and (\ref{diff_eins2}) 
with Eq. (\ref{kol2}),
we find
\bea
&&\Big(\Box_4+6H^2\Big)\big(\phi_2-\phi_3\big)
-\sqrt{3(D-2)\Big(1+\frac{D-6}{\zeta(D-2)}\Big)}H
\big(\phi_2-\phi_3\big){}'
+\big(\phi_2{}-\phi_3{}\big){}''
\nn\\
&=&\frac{f^2}{2} \big(2\phi_3+\alpha\delta\phi\big).
\label{kum2}
\eea
Thus, in contrast to the previous cases of 
the pure gravity and the scalar-tensor theory,
there is the source term in the right-hand side.
Since this source term is proportional to $f^2$,
it is induced if there is the non-zero flux.



\end{document}